\documentclass{ws-procs975x65}

\begin{document}



\title{AN AWESOME HYPOTHESIS FOR DARK ENERGY : THE ABNORMALLY WEIGHTING ENERGY}

\author{ANDR\'E F\"UZFA$^{1,2}$\footnote{F.N.R.S. Postdoctoral Researcher}}
\author{JEAN-MICHEL ALIMI$^2$}

\address{$^1$GAMASCO, Facult\'es Universitaires Notre-Dame de la Paix, Namur, Belgium\\ 
$^2$Laboratory Universe and Theories, CNRS UMR 8102,\\
 Observatoire de Paris-Meudon and Universit\'e Paris VII, France}


\begin{abstract}
We introduce the Abnormally Weighting Energy (AWE) hypothesis in which
dark energy (DE) is presented as a consequence of the violation of the weak equivalence principle (WEP) at cosmological scales by some dark sector.
Indeed, this implies a violation of the strong equivalence principle (SEP) for ordinary matter and consequent cosmic acceleration in the observable frame
as well as variation of the gravitational constant.
The consequent DE mechanism build upon the AWE hypothesis 
(i) does not require a violation of the strong energy condition $p<-\rho c^2/3$, 
(ii) assumes rather non-negligible direct couplings to the gravitational scalar field
(iii) offers a natural convergence mechanism toward general relativity
(iv) accounts fairly for supernovae data from various couplings and equations of state of the dark sector as well as density parameters very close to the
ones of the concordance model $\Lambda CDM$. Finally (v), this AWE mechanism typically ends up with an Einstein-de Sitter expansion regime once
the attractor is reached. 
\end{abstract}

\begin{figure}
\begin{center}
\includegraphics[scale=0.5]{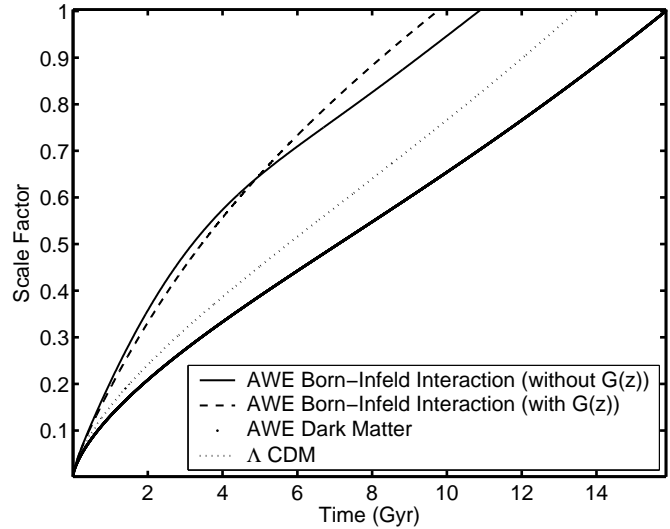}%
\caption{Evolutions of the scale factor in some AWE models and comparison with the concordance model
$\Lambda CDM$ (straight line is the AWE Born-Infeld (BI) gauge interaction\cite{fuzfa} without a variation of $G$ ; the dashed one
is the same model accounting for this variation ($(\tilde{\Omega}_{m,0},\tilde{\Omega}_{BI,0})\approx(0.3,0.7)$); the bold dotted line is an AWE Dark Matter (DM) model\cite{fuzfa2} for which $(\tilde{\Omega}_{m,0},\tilde{\Omega}_{DM,0})\approx(0.04,0.26)$ and dotted line is a 
$(\tilde{\Omega}_{m,0},\tilde{\Omega}_{\Lambda,0})\approx(0.3,0.7)$ $\Lambda CDM$ model)\label{fig1}}
\end{center}
\end{figure}

\bodymatter\bigskip 
The AWE hypothesis can be seen as a generalization of many models aimed to address the DE problem. Let us begin by the
heart of the concordance model: the controversial cosmological constant. 
The theory of gravitation behind the concordance model is the usual Einstein-Hilbert
action of general relativity (GR):
\begin{equation}
\label{eh}
S_{EH}=\frac{1}{2\kappa_*}\int\sqrt{-g_*}d^4x\left\{R^*+2\Lambda\right\}+S_m\left[\psi_m,g^*_{\mu\nu}\right],
\end{equation}
where $g^*_{\mu\nu}$ is the Einstein metric, $\Lambda$ the cosmological constant, $\kappa_*=8\pi G_*$ is the gravitational coupling constant, $R^*$ is the curvature scalar and $\psi_m$ are the matter fields.
The cosmological constant is well-known to lead to intricate fine-tuning and coincidence problems whose possible way-out is to assume a cosmological mechanism for DE which would rule the non-trivial gravitational vacuum. Quintessence 
is problably the most popular of these mechanisms, and is based on
a real 
massive scalar field rolling down some self-interaction potential. 
During a slow-roll phase of this massive scalar field 
(when its kinetic energy is much less than its self-interacting one), 
negative pressures are achieved, producing the desired cosmic acceleration if they violate the strong energy condition (SEC):
$p<-\rho/3$. More precisely, the quintessence mechanism is implemented by the following theory:
\begin{equation}
\label{quint}
S_{Quint}=\frac{1}{2\kappa_*}\int\sqrt{-g_*}d^4x\left\{R^*-2g_*^{\mu\nu}\partial_\mu\varphi\partial_\nu\varphi+V(\varphi)\right\}+S_m\left[\psi_m,g^*_{\mu\nu}\right],
\end{equation}
with $\varphi$ the quintessence scalar field and $V(\varphi)$ its self-interaction potential.
The quintessence mechanism relies strongly on an appropriate choice of this potential whose shapes can be quite sophisticated
in order to reproduce Hubble diagram data.
The cosmological constant $\Lambda$ in (\ref{eh}) corresponds in (\ref{quint}) to a quintessence scalar field frozen ($\partial_\mu\varphi\partial^\mu\varphi\approx 0$)
in a non-vanishing energy state $V(\varphi)> 0$.
However, quintessence is a very discrete form of DE whose interactions with matter are purely gravitational. This can be extended to
non-minimal couplings in order to make this field running physical constants and account for a violation of the equivalence principle (EP). This constitutes the famous tensor-scalar
theories of gravitation
\begin{equation}
\label{st}
S_{TS}=\frac{1}{2\kappa_*}\int\sqrt{-g_*}d^4x\left\{R^*-2g_*^{\mu\nu}\partial_\mu\varphi\partial_\nu\varphi+V(\varphi)\right\}+S_m\left[\psi_m,A_m^2(\varphi)g^*_{\mu\nu}\right],
\end{equation}
where $A_m(\varphi)$ denotes the coupling function to the Einstein metric (quintessence corresponds to $A_m(\varphi)=1$).
However, if we consider that the $70\%$ of missing energy is the whole contribution of such a non-minimally coupled
scalar field, then it might be difficult to match the present tests
of GR without assuming the non-minimal couplings to be \textit{extremely} weak. 
Even worst, as soon as these couplings are non-vanishing, the likely endless domination of a non-minimally coupled scalar
field with quintessence potential might also inescapably lead to a disastrous violation of the EP in the future.\\
\\
The AWE Hypothesis\cite{fuzfa} consists of assuming that DE
does not couple to gravitation in the same way as usual matter (baryons, photons, ...).
This implies a violation of the weak equivalence principle (WEP),
mostly on cosmological scales, where DE dominates the energy content of the Universe. 
This anomalous weight also implies that the related gravitational binding
energy of DE does not generate the same amount of gravity than the one of usual matter, yielding
a violation of the strong equivalence principle (SEP). This results in a running gravitational coupling
constant on cosmological scales whose dynamics will produce cosmic acceleration.
The corresponding action can be written down :
\begin{eqnarray}
\label{action1}
S&=&\frac{1}{2\kappa_*}\int\sqrt{-g_*}d^4x\left\{R^*-2g_*^{\mu\nu}\partial_\mu\varphi\partial_\nu\varphi\right\}
+S_m\left[\psi_m,A_m^2(\varphi)g^*_{\mu\nu}\right]\nonumber\\
&+&S_{awe}\left[\psi_{awe},A_{awe}^2(\varphi)g^*_{\mu\nu}\right],
\end{eqnarray}
with $S_{awe}$ is the action for the AWE
sector with fields $\psi_{awe}$ and $S_m$ is the usual matter
sector with matter fields $\psi_m$; $A_{awe}(\varphi)$ and $A_m(\varphi)$ being the coupling functions 
to the metric $g^*_{\mu\nu}$ for the AWE and matter sectors respectively. Usual tensor-scalar theories (\ref{st}) correspond
to a universal coupling $A_{awe}(\varphi)=A_m(\varphi)$.
However, the action (\ref{action1}) is written in the so-called "\textit{Einstein frame}" 
of the physical degrees of freedom while cosmology is expressed in the observable frame $\tilde{g}_{\mu\nu}=A_m^2(\varphi)g^*_{\mu\nu}$.
This subtelty will allow to produce cosmic acceleration without violation of SEC through the acceleration of $\varphi$ in $\tilde{a}=A_m(\varphi)a_*$.
The AWE mechanism is based on a competition between the non-minimal couplings of the two material sectors to rule the EP. 
This competition is similar to 
the dynamics of the so-called chameleon fields \cite{chameleon} between the self-interaction term and the non-minimal coupling.
The competition ends up with the stabilization of $\varphi$ on a attractor, leaving the corresponding gravitational theory asymptotically similar to
GR. During some time before the stabilization, both transient cosmic acceleration and variation of $G$ appear allowing to explain Hubble diagram of type Ia supernovae. This mechanism has been implemented in the case of an AWE Born-Infeld gauge interaction\cite{fuzfa} but also works with a pressureless fluid\cite{fuzfa2}. Figure \ref{fig1}
illustrates the behavior of the scale factor in different AWE cosmologies fitting the Hubble diagram of type Ia supernovae. The AWE hypothesis 
exhibits key features for DE (see the abstract) and we are confident that it will surely constitute a seducing alternative to the cosmological constant.

\vfill

\end{document}